\documentclass[alpha-refs]{wiley-article}

\usepackage{graphicx}
\usepackage[space]{grffile}
\usepackage{latexsym}
\usepackage{textcomp}
\usepackage{longtable}
\usepackage{tabulary}
\usepackage{booktabs,array,multirow}
\usepackage{amsfonts,amsmath,amssymb}
\usepackage{natbib}
\usepackage{moreverb,url}
\usepackage{bm}
\usepackage{adjustbox}
\usepackage[colorlinks,bookmarksopen,bookmarksnumbered,citecolor=blue,urlcolor=blue,linkcolor=blue]{hyperref}
\usepackage{fancyhdr,rotating}
\usepackage{longtable}
\usepackage{pdflscape}
\usepackage{color,epsfig}
\hypersetup{colorlinks=false,pdfborder={0 0 0}}
\usepackage{etoolbox}

\newcommand{\bfa}{\bm{a}}
\newcommand{\bff}{\bm{f}}
\newcommand{\bfg}{\bm{g}}
\newcommand{\bfh}{\bm{h}}

\newcommand{\bfq}{\bm{q}}
\newcommand{\bfs}{\bm{s}}

\newcommand{\bfx}{\bm{x}}

\newcommand{\bfbeta}{\bm{\beta}}
\newcommand{\bfeta}{\bm{\eta}}
\newcommand{\bfepsilon}{\bm{\epsilon}}

\newcommand{\bfzeta}{\bm{\zeta}}
\newcommand{\bfpsi}{\bm{\psi}}

\newcommand{\bfPhi}{\bm{\Phi}}

\newcommand{\bfSigma}{\bm{\Sigma}}
\newcommand{\bfOmega}{\bm{\Omega}}

\newcommand{\rmd}{\mathrm{d}}

\newcommand{\bfA}{\bm{A}}
\newcommand{\bfC}{\bm{C}}

\newcommand{\bfQ}{\bm{Q}}
\newcommand{\bfS}{\bm{S}}
\newcommand{\bfT}{\bm{T}}
\newcommand{\bfU}{\bm{U}}
\newcommand{\bfW}{\bm{W}}
\newcommand{\bfX}{\bm{X}}
\newcommand{\bfZ}{\bm{Z}}

\newcommand{\bbE}{\mathbb{E}}
\newcommand{\bbI}{\mathbb{I}}
\newcommand{\bbP}{\mathbb{P}}
\newcommand{\bbR}{\mathbb{R}}

\newcommand{\calN}{\mathcal{N}}

\newcommand{\calS}{\mathcal{S}}
\newcommand{\calU}{\mathcal{U}}

\newif\iflatexml\latexmlfalse

\AtBeginDocument{\DeclareGraphicsExtensions{.pdf,.PDF,.eps,.EPS,.png,.PNG,.tif,.TIF,.jpg,.JPG,.jpeg,.JPEG}}

\usepackage[utf8]{inputenc}
\usepackage[english]{babel}

\usepackage{siunitx}

\iflatexml

\else

\paperfield{Statistica Neerlandica}
\corraddress{Xinyuan Chen, Department of Mathematics and Statistics, Mississippi State University, Mississippi State, MS, 39762, USA}
\corremail{xchen@math.msstate.edu}
\fundinginfo{National Center for Advancing Translational Science (NCATS), National Institutes of Health (NIH), Grant Number: UL1 TR0001863; Patient Centered Outcomes Research Institute (PCORI), National Institute on Aging (NIA) at NIH, Grant Numbers: 5U01AG048270; P30AG21342.}
\fi

\papertype{Original Article}

\title{Competing risks regression for clustered survival data via the marginal additive subdistribution hazards model}

\author[1]{Xinyuan Chen}

\affil[1]{Department of Mathematics and Statistics, Mississippi State University, Mississippi State, MS, USA}

\author[2,3]{Denise Esserman}

\affil[2]{Department of Biostatistics, Yale School of Public Health, New Haven, CT, USA}

\author[2,3,4]{Fan Li}

\affil[3]{Yale Center for Analytical Sciences, Yale School of Public Health, New Haven, CT, USA}

\affil[4]{Center for Methods in Implementation and Preventive Science, Yale School of Public Health, New Haven, CT, USA}

\runningauthor{Xinyuan Chen et. al.}

\begin{document}

\maketitle
\selectlanguage{english}
\begin{abstract}
A population-averaged additive subdistribution hazards model is proposed to assess the marginal effects of covariates on the cumulative incidence function and to analyze correlated failure time data subject to competing risks. This approach extends the population-averaged additive hazards model by accommodating potentially dependent censoring due to competing events other than the event of interest. Assuming an independent working correlation structure, an estimating equations approach is outlined to estimate the regression coefficients and a new sandwich variance estimator is proposed. The proposed sandwich variance estimator accounts for both the correlations between failure times and between the censoring times, and is robust to misspecification of the unknown dependency structure within each cluster. We further develop goodness-of-fit tests to assess the adequacy of the additive structure of the subdistribution hazards for the overall model and each covariate.  Simulation studies are conducted to investigate the performance of the proposed methods in finite samples. We illustrate our methods using data from the STrategies to Reduce Injuries and Develop confidence in Elders (STRIDE) trial.

\end{abstract}%

\section{Introduction}\label{sec:intro}

Competing risks data commonly arise in randomized trials and observational studies when the occurrence of the event of primary interest is precluded by a different event. Without assuming independence between events of different causes, censoring participants who fail from the competing event can often lead to an overestimation of the cumulative probabilities of the primary event \citep{lau2009competing}. To study the effects of covariates on the cumulative probabilities of a particular cause, \citet{Fine1999} proposed a proportional subdistribution hazards model, which has a one-to-one correspondence to the cumulative incidence function. The subdistribution hazards model treats the competing causes differently from independent censoring variables and has become one of the mainstream regression approaches for analyzing competing risks data. In cluster randomized trials or familial studies, however, the classic Fine and Gray model may not be directly applicable as the assumption of independent participants may no longer be valid; for example, participants sharing the same provider in the same clinic, or members of the same family sharing the same unobserved risk factors. Such clustered survival data are referred to as \emph{clustered competing risks}, where cluster contributes to the dependence among the observations collected for the cluster members and competing risks contributes to the dependence across causes of failures \citep{diao2013clustered,Zhou2012}. The unknown but potentially complex within-cluster dependency structure among the failure observations necessitates appropriate regression methods that account for the correlations in a robust manner to enable valid inference for the covariate effects on the cumulative incidence function of the primary event.

Conditional and marginal models represent two modeling strategies for cumulative incidence regression of clustered competing risks data. Random effects are used to account for correlations among failure observations in the conditional subdistribution hazards model and the fixed-effects parameters often have a cluster-specific interpretation. For example, \citet{Katsahian2006} and \citet{katsahian2011estimating} developed the frailty model for the subdistribution hazards when competing risks arise in clustered designs. \citet{dixon2011competing} and \citet{dixon2012applying} considered alternative specifications of the frailty terms to allow for marginal interpretation of regression coefficients. Similar frailty \citep{Christian2016,Ha2016, Lee2016,Rueten2019} and copula models \citep{Emura2020} have also been developed for the cause-specific hazards regression. While these models are flexible in so far as explicitly modeling the heterogeneity across clusters, valid inference for fixed-effects regression parameters necessarily depends on the correct specification of the frailty distribution or copula. In contrast, marginal models specify the covariate effects that are averaged across the population of clusters and do not require the specification of the unobserved frailties \citep{liang1986longitudinal}. \citet{Zhou2012} proposed a marginal subdistribution model and an estimation strategy assuming an independence working correlation structure. A sandwich variance estimator was developed to account for the unknown within-cluster dependency structures. Furthermore, the marginal subdistribution hazards model is specified unconditionally on the latent random effects, with the regression coefficients bearing a population-averaged interpretation. With the sandwich variance estimator, inference for the marginal regression models is generally robust to assumptions of the within-cluster correlations between failure times and those between censoring times. 


While the marginal subdistribution hazards model can be appealing for analyzing clustered competing risks data, the development in \citet{Zhou2012} relies on the proportional subdistribution hazards assumption, which may not always hold in practice and violation of which can lead to bias. Therefore, it remains of interest to investigate complementary approaches to modeling the cumulative incidence function with clustered competing risks data. In this article, we develop an additive subdistribution hazards model for clustered competing risks regression that does not rely on the proportional subdistribution hazards assumption. The additive subdistribution hazards model is akin to the additive risk model in the absence of competing risks \citep{buckley1984additive,Lin1994,yin2004additive} as well as the additive cause-specific hazards model with competing risks \citep{shen1999confidence}, but is different from them as the additive structure is assumed for the subdistribution hazards, which has a one-to-one correspondence to the cumulative incidence function. Assuming an independence working correlation model, we estimate model parameters through the generalized estimating equations (GEE) framework, similar to the estimation approach in \citet{Li2017} and \citet{Sun2006} in the absence of clustering, and to \citet{Wogu2023} for case-cohort competing risks data in the absence of clustering. We further develop a new sandwich variance estimator to properly account for clustering of both the failure and censoring observations. In addition, methods for model checking and goodness-of-fit tests can aid in the credibility of competing risks regression analysis. In the absence of clustering, \citet{Sun2006} developed a set of model checking tests for the combined additive-multiplicative subdistribution hazards model and \citet{Li2015} developed goodness-of-fit tests to assess the proportional subdistribution hazards assumption for the Fine and Gray model. However, those tests are not directly applicable to clustered survival data and competing risks due to the presence of within-cluster correlations. We propose a class of model checking and goodness-of-fit tests based on the weighted martingale residual process from the proposed marginal additive subdistribution hazards model, extending the approach in \citet{Yin2007} to the analysis of clustered competing risks data. The class of tests can detect different aspects of model misspecification including the assessment of the additive structure in the subdistribution hazards as well as the functional form of covariates. Theoretical properties of the proposed estimators and testing procedures are established, and their empirical performance is studied via simulations and a real data application.

The remainder of the article is organized as follows. Section \ref{sec:model} introduces the marginal additive subdistribution hazards model and Section \ref{sec:estimation} develops the details for estimation and inference for its model parameters. 
In Section \ref{sec:checking}, we propose a class of model checking procedures for assessing key aspects of model misspecification. Section \ref{sec:sim} presents results from simulation studies evaluating the finite-sample performance of the proposed model and the model checking procedure. An illustrative analysis of a cluster randomized clinical trial is presented in Section \ref{sec:app}. 
Section \ref{sec:dis} concludes with a discussion. 

\section{Marginal additive subdistribution hazards model} \label{sec:model}

We consider a study with clustered time-to-event outcomes collected from $n$ independent clusters. Define $T_{ij}$ as the failure time of the $j$th participant ($j=1,\dots,m_i$) in the $i$th cluster ($i=1,\dots,n$). In the presence of competing risks, we let $\epsilon_{ij}\in\{1,\dots,K\}$ denote different causes of failure and $\bfX_{ij}(t)$ be a $p$-vector of bounded external covariates, which can be time-dependent \citep{Austin2019review}. Defining the population-averaged (marginal) cumulative incidence function (CIF) for the failure from cause $k$ adjusted for covariates as $F_k(t;\bfX) \equiv \bbP\{T\leq t,\epsilon = k|\bfX(t)\}$, we can write the corresponding marginal subdistribution hazard as $\lambda_k(t;\bfX)=-\rmd\log\{1-F_k(t;\bfX)\}/\rmd t$. The marginal additive subdistribution hazards model corresponding to cause $k$ for clustered competing risks data is then given by
\begin{align} \label{eq:model-def}
	\rmd\Lambda_{ijk}(t) = \rmd\Lambda_{0k}(t) + \bfX_{ij}(t)'\bfbeta_k\rmd t,
\end{align}
where $\Lambda_{0k}(t)$ is the unspecified baseline subdistribution hazards, $\bfbeta_k$ is the associated $p$-vector of regression coefficients, and the cumulative subdistribution hazard function for each participant is $\Lambda_{ijk}(t) = \Lambda_{0k}(t) + \int_0^t \bfX_{ij}(u)'\bfbeta_k\rmd u$.  Of note, the marginal additive subdistribution hazards model for cause $k$ does not require assumptions regarding the CIF for other causes. Also, we note that model \eqref{eq:model-def} does not assume proportional subdistribution hazards and serves as a complementary approach to model the CIF when the assumption of proportional subdistribution hazards is no longer considered plausible.

Assuming in practice, failure time $T_{ij}$ is right-censored,  we define $C_{ij}$ as the censoring time, and we observe $\{Z_{ij}=T_{ij}\wedge C_{ij},\zeta_{ij}=\bbI(T_{ij}\leq C_{ij})\epsilon_{ij},\bfX_{ij}(t)\}$ for each participant. Let $\bfT_i=(T_{i1},\ldots,T_{im_i})$, $\bfepsilon_i=(\epsilon_{i1},\ldots,\epsilon_{im_i})$, $\bfX_i(t)=(\bfX_{i1}(t),\ldots,\bfX_{im_i}(t))$, and $\bfC_i=(C_{i1},\ldots,C_{im_i})$. We assume $\{(\bfT_i,\bfepsilon_i,\bfX_i,\bfC_i,m_i), i=1,\dots,n\}$ are independent and identically distributed (i.i.d.). We also assume $(\bfT_i, \bfepsilon_i)$ and $\bfC_i$ are independent conditional on $(\bfX_i, m_i)$ for each cluster $i$. In each cluster $i$, however, the components of $(\bfT_i, \bfepsilon_i)$ can be arbitrarily correlated given $(\bfX_i, m_i)$, and similarly, the components of $\bfC_i$ can also be arbitrarily correlated given $(\bfX_i, m_i)$. We further denote $\bfZ_i=(Z_{i1},\dots,Z_{im_i})$ and $\bfzeta_i=(\zeta_{i1},\dots,\zeta_{im_i})$ as the collection of observed survival times and the observed event indicators within each cluster, respectively. The observed data for each cluster, $(\bfZ_i, \bfzeta_i, \bfX_i, m_i)$, are i.i.d. due to the aforementioned assumptions.

\section{Estimation and inference for model parameters}\label{sec:estimation}

We present estimation procedures for the marginal additive subdistribution hazards model, starting with the case of censoring complete (CC) data, where failure time $T$ is right-censored but the potential censoring time $C$ is always observed. The estimation procedure for the CC data paves the way for the more common scenario with right-censored data.

\subsection{Censoring complete data}\label{sec:CCdata}

We start with the case of complete censoring (CC) where the failure time $T$ is right-censored and censoring occurs only due to administrative loss to follow-up, independent of covariates. Hereafter, we assume that there exists a maximum follow-up time $\tau<\infty$ such that $\bbP(T_{ij}>\tau) > 0$, and $\bbP(C_{ij}=\tau) = \bbP(C_{ij}\geq\tau) > 0$ for participant $j$ in cluster $i$. Suppose cause of failure $k$ is of particular interest, then let $N_{ij}^k(t) = \bbI (T_{ij}\leq t,\zeta_{ij}=k)$ and $Y_{ij}^k(t)=1-N_{ij}^k(t-)$ denote the counting process and the risk process, respectively. In particular, for cumulative incidence regression, the risk set involves those who have not failed from any cause as well as those who have previously failed from other causes not of primary interest. For CC data, since the censoring time is assumed to be observable, the risk process is modified to $Y_{ij}^{k*}(t) = \bbI(C_{ij}>t)Y_{ij}^k(t)$. To proceed, we establish the following notation:
\begin{align*}
	\bfS^{(r)}(t) &= n^{-1}\sum_{i=1}^n\sum_{j=1}^{m_i}\bbI(C_{ij}>t)Y_{ij}^k(t)\bfX_{ij}(t)^{\otimes r},~~~\bfs^{(r)}(t) = \lim_{n\rightarrow\infty}\bfS^{(r)}(t), ~~~ r=0,1,\displaybreak[0]\\
	\bar{\bfX}(t) &= \bfS^{(1)}(t)/S^{(0)}(t), ~~~ \bar{\bfx}(t) = \bfs^{(1)}(t)/s^{(0)}(t),\displaybreak[0]\\
	\bar{\bfA}(\tau) &= n^{-1}\sum_{i=1}^n\sum_{j=1}^{m_i}\int_0^\tau\bbI(C_{ij}>t)Y_{ij}^k(t)\left\{\bfX_{ij}(t)-\bar{\bfX}(t)\right\}^{\otimes2}\rmd t,
\end{align*}
where $\bfa^{\otimes0}=1$, $\bfa^{\otimes1}=\bfa$, and $\bfa^{\otimes2}=\bfa\bfa'$. We further assume the following regularity conditions hold: 
\begin{itemize}
	\item[(R1).] The baseline subdistribution hazards satisfy $\int_0^\tau\rmd\Lambda_{0k}(t)<\infty$; 
	\item[(R2).] $\bfX_{ij}(t)$ has bounded total variations on $\bbR^p$ almost surely $\forall~i$ and $j$;
	\item[(R3).] $\bar{\bfA}(\tau)$ converges to a positive definite matrix $\bfA^*(\tau)$.
\end{itemize}
We can then estimate $\bfbeta_k$ by solving the following set of estimating equations under an independence working assumption:
\begin{align} \label{eq:cc-score}
	\bfU(\bfbeta_k) = \sum_{i=1}^n\sum_{j=1}^{m_i}\int_0^\tau\left\{\bfX_{ij}(t)-\bar{\bfX}(t)\right\}\bbI(C_{ij}>t)\rmd M_{ij}^k(\bfbeta_k,t),
\end{align}
where
\begin{align} \label{eq:cc-mart}
	M_{ij}^k(\bfbeta_k,t) = N_{ij}^k(t)- \int_0^t Y_{ij}^k(u)\left\{\rmd \Lambda_{0k}(u)+\bfX_{ij}(u)'\bfbeta_k\right\}\rmd u
\end{align}
is a martingale for the marginal data filtration $\mathcal{F}_{ij}^k(t)=\{N_{ij}^k(u),Y_{ij}^k(u),\allowbreak Y_{ij}^k(u)\times\allowbreak\bfX_{ij}(u);\allowbreak u\leq t\}$ generated from each participant $j$ of cluster $i$. However, due to within-cluster correlations, $M_{ij}^k(\bfbeta_k,t)$ is not a martingale for the joint filtration generated by all the failure, censoring, and covariate information across clusters up to time $t$. Setting $\bfU(\bfbeta_k)=\bm{0}$, we obtain the ordinary least squares estimator for $\bfbeta_k$:
\begin{align*} 
	\hat{\bfbeta}_k =& \left[\sum_{i=1}^n\sum_{j=1}^{m_i}\int_0^\tau\bbI(C_{ij}>t)Y_{ij}^k(t)\left\{\bfX_{ij}(t)-\bar{\bfX}(t)\right\}^{\otimes2}\rmd t\right]^{-1}
	\left[\sum_{i=1}^n\sum_{j=1}^{m_i}\int_0^\tau\left\{\bfX_{ij}(t)-\bar{\bfX}(t)\right\}\bbI(C_{ij}>t)\rmd N_{ij}^k(t)\right].
\end{align*}
We also estimate the baseline subdistribution hazards by
\begin{align*} 
	\hat{\Lambda}_{0k}(t) = \int_0^t \frac{\sum_{i=1}^n\sum_{j=1}^{m_i}\bbI(C_{ij}>u)Y_{ij}^k(u)\left\{\rmd N_{ij}^k(u)-\bfX_{ij}(u)'\hat{\bfbeta}_k\rmd u\right\}}{\sum_{i=1}^n\sum_{j=1}^{m_i}\bbI(C_{ij}>t)Y_{ij}^k(u)}.
\end{align*}
For CC data, $\hat{\bfbeta}_k$ and $\hat{\Lambda}_{0k}(t)$ are natural extensions of the estimators developed for the marginal additive hazards model in the absence of competing risks \citep{yin2004additive}, and thus have similar large-sample properties. Proofs for consistency and asymptotic normality of these estimators with the CC data are also similar to and relatively simpler than those with right-censored data developed in Section \ref{sec:RCdata}; technical details for the CC data are given in Web Appendix A2. 

\subsection{Right-censored data}\label{sec:RCdata}

When the data are right-censored, we modify the least squares estimating equation \eqref{eq:cc-score} by using the inverse probability of censoring weight (IPCW) approach \citep{Robins1992} for the marginal proportional subdistribution hazards model. We assume that the censoring time $C_{ij}$ is independent of the covariates $\bfX_{ij}(t)$ for simplicity; extensions to covariate-dependent censoring mechanisms are possible along the lines of \citet{he2016proportional}. Specifically, the vital status for the $j$th participant in the $i$th cluster at time $t$ is denoted as $r_{ij}(t)=\bbI(C_{ij}\geq T_{ij}\wedge t)$, and $G(t)=\bbP(C_{ij}\geq t)$ is the survival function of censoring times. We then define the time-dependent IPCW $\hat{\omega}_{ij}(t)=r_{ij}(t)\hat{G}(t)/\hat{G}(Z_{ij}\wedge t)$, where $\hat{G}(t)$ is the Kaplan-Meier (KM) estimate of $G(t)$ based on the data $\{(Z_{ij}, 1-\Delta_{ij}), j=1,\dots,m_i,i=1,\dots,n\}$, with $\Delta_{ij} = \bbI(T_{ij}\leq C_{ij})$ \citep{Fine1999}. To proceed, we define the following notation based on the IPCWs:
\begin{align}
	\hat{\bfS}^{(r)}(t) &= n^{-1}\sum_{i=1}^n\sum_{j=1}^{m_i}\hat{\omega}_{ij}(t)Y_{ij}^k(t)\bfX_{ij}(t)^{\otimes r}, ~~~\tilde{\bfs}^{(r)}(t) = \lim_{n\rightarrow\infty}\hat{\bfS}^{(r)}(t), ~~~ r=0,1,\displaybreak[0]\nonumber\\
	\hat{\bfX}(t) &= \hat{\bfS}^{(1)}(t)/\hat{S}^{(0)}(t), ~~~ \tilde{\bfx}(t) = \tilde{\bfs}^{(1)}(t)/\tilde{s}^{(0)}(t),\displaybreak[0]\nonumber\\
	\hat{\bfA}(\tau) &= n^{-1}\sum_{i=1}^n\sum_{j=1}^{m_i}\int_0^\tau\hat{\omega}_{ij}(t)Y_{ij}^k(t)\left\{\bfX_{ij}(t)-\hat{\bfX}(t)\right\}^{\otimes2}\rmd t.\label{eq:Atau}
\end{align}
Assuming regularity conditions (R1) and (R2) in Section \ref{sec:CCdata} hold but replacing (R3) with 
\begin{itemize}
	\item[(R4).] $\hat{\bfA}(\tau)$ converges to a positive definite matrix $\bfA(\tau)$,
\end{itemize}
we can estimate $\bfbeta_k$ via the following IPCW estimating equations under the working independence assumption \citep{Li2017},
\begin{align} \label{eq:rc-score}
	\bfU(\bfbeta_k) = \sum_{i=1}^n\sum_{j=1}^{m_i}\int_0^\tau\left\{\bfX_{ij}(t)-\hat{\bfX}(t)\right\}\hat{\omega}_{ij}(t)\rmd M_{ij}^k(\bfbeta_k,t),
\end{align}
with $M_{ij}(\bfbeta_k,t)$ as the martingale defined in (\ref{eq:cc-mart}).
Setting $\bfU(\bfbeta_k)=\bm{0}$, we obtain the weighted least squares estimator for ${\bfbeta}_k$
\begin{align} \label{eq:rc-beta-est}
	\hat{\bfbeta}_k =& \left[\sum_{i=1}^n\sum_{j=1}^{m_i}\int_0^\tau\hat{\omega}_{ij}(t)Y_{ij}^k(t)\left\{\bfX_{ij}(t)-\hat{\bfX}(t)\right\}^{\otimes2}\rmd t\right]^{-1}
	\left[\sum_{i=1}^n\sum_{j=1}^{m_i}\int_0^\tau\left\{\bfX_{ij}(t)-\hat{\bfX}(t)\right\}\hat{\omega}_{ij}(t)\rmd N_{ij}^k(t)\right].
\end{align}
By weighting the risk process, $\Lambda_{0k}(t)$ can be estimated by
\begin{align} \label{eq:rc-base-est}
	\hat{\Lambda}_{0k}(t) = \int_0^t \frac{\sum_{i=1}^n\sum_{j=1}^{m_i}\hat{\omega}_{ij}(u)Y_{ij}^k(u)\left\{\rmd N_{ij}^k(u)-\bfX_{ij}(u)'\hat{\bfbeta}_k\rmd u\right\}}{\sum_{i=1}^n\sum_{j=1}^{m_i}\hat{\omega}_{ij}(u)Y_{ij}^k(u)}.
\end{align}
Let $\bfbeta_{k,0}$ and $\Lambda_{0k,0}(t)$ denote the true values for $\bfbeta_k$ and $\Lambda_{0k}(t)$, respectively. To enable statistical inference on $\bfbeta_k$ and $\Lambda_{0k}(t)$ with clustered competing risks data, we first establish the following large-sample results. 
\begin{theorem}\label{thm:es-1}
	Under regularity conditions (R1), (R2), and (R4), $\hat{\bfbeta}_k$ is consistent and asymptotically normal, with $\sqrt{n}(\hat{\bfbeta}_k-\bfbeta_{k,0})$ converging in distribution to a mean-zero Gaussian random variate. Furthermore, $\hat{\Lambda}_{0k}(t)$ is uniformly consistent, with $\sqrt{n}\{\hat{\Lambda}_{0k}(t)-\Lambda_{0k,0}(t)\}$ converging weakly to a mean-zero Gaussian process in $l^{\infty}[0,\tau]$.
\end{theorem}
The proof of Theorem \ref{thm:es-1} is given in Web Appendix A3. One key step in deriving the asymptotic results is to recognize that the KM estimator $\hat{G}(t)$ is consistent even when the censoring times are correlated within each cluster. Intuitively, one can consider $\hat{G}(t)$ as the solution to an estimating equation with an independence working assumption, and therefore establish the consistency of the KM estimator with clustered censoring observations \citep{Zhou2012}.

The covariance matrix of the asymptotic distribution of $\sqrt{n}(\hat{\bfbeta}_k-\bfbeta_{k,0})$, denoted by $\bfSigma_{\bfbeta_k}$, has a sandwich form $\bfSigma_{\bfbeta_k} = \bfA^{-1}(\tau)\bfOmega\bfA^{-1}(\tau)$, where $\bfOmega$ is the variance of the asymptotic normal distribution of the root-$n$ scaled estimating equations, $n^{-1/2}\bfU(\bfbeta_{k})$ evaluated at $\bfbeta_{k,0}$. The expression of $\bfOmega$ requires the specification of a few more quantities. Let $N_{ij}^c(t) = \bbI(T_{ij} \leq t,\Delta_{ij}=0)$ denote the counting process of the censoring times and $Y_{ij}^c(t)=1-N_{ij}^c(t-)$ denote the associated risk process. It follows that $M_{ij}^c(t) = N_{ij}^c(t) - \int_0^t Y_{ij}^c(u)\rmd\Lambda_0^c(u)$ is the martingale associated with marginal filtration generated by information from the $j$th participant in the $i$th cluster, and $\Lambda_0^c(t)$ is the common cumulative hazards function of the censoring variable. This notation allows us to write $\bfeta_{ij} = \int_0^\tau \{\bfX_{ij}(t)-\tilde{\bfx}(t)\}\omega_{ij}(t)\rmd M_{ij}^k(\bfbeta_{k},t)$ and $\bfpsi_{ij} = \int_0^\tau {\bfq(u)}{\pi^{-1}(u)}\rmd M_{ij}^c(u)$, where
	\begin{align*}
		&\bfq(u) = -\lim_{n\rightarrow\infty}n^{-1}\sum_{i=1}^n\sum_{j=1}^{m_i}\int_0^\tau\left\{\bfX_{ij}(t)-\tilde{\bfx}(t)\right\}\omega_{ij}(t)\bbI(Z_{ij}<u\leq t)\rmd M_{ij}^k(\bfbeta_{k},t),\displaybreak[0]\\
		&\pi(u)=\lim_{n\rightarrow\infty}n^{-1}\sum_{i=1}^n\sum_{j=1}^{m_i}Y_{ij}^c(u).
	\end{align*}
	We then obtain $\bfOmega=\bbE\{(\bfeta_{i\cdot}+\bfpsi_{i\cdot})^{\otimes2}\}$ where $\bfeta_{i\cdot}=\sum_{j=1}^{m_i}\bfeta_{ij}$ and $\bfpsi_{i\cdot}=\sum_{j=1}^{m_i}\bfpsi_{ij}$. 
The form of the covariance matrix suggests a consistent sandwich variance estimator for $\bfbeta_k$ as $\hat{\bfSigma}_{\bfbeta_k} = \hat{\bfA}^{-1}(\tau)\hat{\bfOmega}\hat{\bfA}^{-1}(\tau)$, where $\hat{\bfA}(\tau)$ is given in \eqref{eq:Atau} and the ``meat'' of the sandwich variance estimator is an empirical variance estimator obtained by averaging the contribution from each cluster $\hat{\bfOmega} = n^{-1}\sum_{i=1}^n(\hat{\bfeta}_{i\cdot}+\hat{\bfpsi}_{i\cdot})^{\otimes2}$, where
\begin{align*}
	&\hat{\bfeta}_{i\cdot} = \sum_{j=1}^{m_i}\int_0^\tau \left\{\bfX_{ij}(t)-\hat{\bfX}(t)\right\}\hat{\omega}_{ij}(t)\rmd \hat{M}_{ij}^k(\hat{\bfbeta}_k,t), ~~~\hat{\bfpsi}_{i\cdot} = \sum_{j=1}^{m_i}\int_0^\tau \frac{\hat{\bfq}(t)}{\hat{\pi}(t)}\rmd \hat{M}_{ij}^c(t), \displaybreak[0]\\
	&\hat{\bfq}(u) = -n^{-1}\sum_{i=1}^n\sum_{j=1}^{m_i}\int_0^\tau\left\{\bfX_{ij}(t)-\hat{\bfX}(t)\right\}\hat{\omega}_{ij}(t)\bbI(Z_{ij}<u\leq t)\rmd \hat{M}_{ij}^k(\hat{\bfbeta}_k,t), \displaybreak[0]\\
	&\rmd \hat{M}_{ij}^k(\hat{\bfbeta}_k,t) = \rmd N_{ij}^k(t) - Y_{ij}^k(t)\left\{\rmd \hat{\Lambda}_{0k}(t) + \bfX_{ij}(t)'\hat{\bfbeta}_k\rmd t\right\}, \displaybreak[0]\\
	&\rmd \hat{M}_{ij}^c(t) = \rmd N_{ij}^c(t) - Y_{ij}^c(t) \rmd\hat{\Lambda}_0^c(t), \displaybreak[0]\\
	&\hat{\pi}(u) = n^{-1}\sum_{i=1}^n\sum_{j=1}^{m_i}Y_{ij}^c(u), ~~~
	\hat{\Lambda}_0^c(t) = \int_0^\tau\frac{\sum_{i=1}^n\sum_{j=1}^{m_i}\rmd N_{ij}^c(t)}{\sum_{i=1}^n\sum_{j=1}^{m_i}Y_{ij}^c(t)}.
\end{align*}

With $\hat{\bfbeta}_k$ and $\hat{\Lambda}_{0k}(\cdot)$, one can estimate the CIF due to cause $k$ for each participant by $\hat{F}_k(t,\bfX_{ij})=1-\exp\{-\hat{\Lambda}_{0k}(t)-\int_0^t \bfX_{ij}(u)'\hat{\bfbeta}_k\rmd u\}$. The large-sample properties developed in Theorem \ref{thm:es-1} further allow one to show that $\sqrt{n}\{\hat{F}_k(t,\bfX_{ij})-F_{k,0}(t,\bfX_{ij})\}$ converges weakly to a mean-zero Gaussian process on $l^\infty[0,\tau]$. The limiting covariance process can then be used to develop a consistent pointwise variance estimator for the estimated CIF, the details of which are given in Web Appendix A3. 

\section{Model checking via goodness-of-fit tests} \label{sec:checking}

To aid in the credibility of analysis using the marginal additive subdistribution hazards model, we further develop objective model checking and goodness-of-fit testing procedures for clustered competing risks data, extending the approach of \citet{Yin2007} developed in the absence of competing risks. Focusing on the case with right-censored data, we propose a general class of tests based on the weighted cumulative sum of martingale transforms over all cluster observations. The proposed testing procedures can be used to assess the additive structure of any specific group of covariates specified in the subditribution hazards model and can be adapted to assess whether other aspects of model misspecification exist (i.e., the functional form of each covariate) \citep{Yin2007,Feng2022}.

To proceed, we define the martingale residual as
\begin{align*} 
	\hat{M}_{ij}^k(\hat{\bfbeta}_k,t) = N_{ij}^k(t)- \int_0^t Y_{ij}^k(u)\left\{\rmd \hat{\Lambda}_{0k}(u)+\bfX_{ij}(u)'\hat{\bfbeta}_k\rmd u\right\}, 
\end{align*} 
which, intuitively, can be viewed as the difference at time $t$ between the observed and expected number of failures due to cause $k$ for the $j$th participant in the $i$th cluster. We can then define a class of cumulative sums of the residuals at time $t$ with IPCW as
\begin{align} \label{eq:cum-sum-res}
	&\bfW(t,\bfx)
	= \sum_{i=1}^n\sum_{j=1}^{m_i}\int_0^t \hat{\omega}_{ij}(u)\bff\left\{\bfX_{ij}(u)\right\}\bbI\left\{\bfX_{ij}(u)\leq\bfx\right\}\rmd\hat{M}_{ij}^k(\hat{\bfbeta}_k,u),
\end{align}
where $\bff(\cdot)$ is a specified vector-valued bounded function, and $\bbI\{\bfX_{ij}(u)\leq\bfx\}=\bbI\{X_{ij1}(u)\leq x_1,\dots,X_{ijp}(u)\leq x_p\}$.  The stochastic process $\bfW(t,\bfx)$ incorporates several specific tests for different aspects of model misspecification, and will be further elaborated after Theorem \ref{thm:mc-1}. For subsequent presentations, we further define
\begin{align*} 
	\hat{\bfg}(t,\bfx) = \frac{\sum_{i=1}^{n}\sum_{j=1}^{m_i}\hat{\omega}_{ij}(t)\bff\left\{\bfX_{ij}(t)\right\}\bbI\left\{\bfX_{ij}(t)\leq\bfx\right\}Y_{ij}^k(t)}{\sum_{i=1}^{n}\sum_{j=1}^{m_i}\hat{\omega}_{ij}(t)Y_{ij}^k(t)}
\end{align*}
and $\hat{\bfh}(t,\bfx)
= \sum_{i=1}^n\sum_{j=1}^{m_i}\int_0^t\hat{\omega}_{ij}(u)Y_{ij}^k(u)\bff\{\bfX_{ij}(u)\}\bbI\{\bfX_{ij}(u)\leq\bfx\}\{\bfX_{ij}(u)-\hat{\bfX}(u)\}\rmd u$, with $\bfg(t,\bfx)=\lim_{n\rightarrow\infty}\hat{\bfg}(t,\bfx)$ and $\bfh(t,\bfx)=\lim_{n\rightarrow\infty}\hat{\bfh}(t,\bfx)$. We write
\begin{align*} 
	\hat{\bfA}(t) = n^{-1}\sum_{i=1}^n\sum_{j=1}^{m_i}\int_0^t\hat{\omega}_{ij}(u)Y_{ij}^k(u)\left\{\bfX_{ij}(u)-\hat{\bfX}(u)\right\}^{\otimes2}\rmd u
\end{align*}
which, under a regularity condition similar to (R4), converges uniformly to a positive definite matrix $\bfA(t)$. Invoking the regularity conditions outlined in Section \ref{sec:RCdata}, we show in Web Appendix A4 that $n^{-1/2}\bfW(t,\bfx)$ converges weakly to $n^{-1/2}\tilde{\bfW}(t,\bfx)= \sum_{i=1}^n \bfQ_i(t,\bfx)$,  a mean-zero Gaussian process with the covariance function between $(t,\bfx)$ and $(t^*,\bfx^*)$ given by $\bbE\{\bfQ_i(t,\bfx)\bfQ_i(t^*,\bfx^*)'\}$, where
\begin{align*}
	\bfQ_i(t,\bfx) =& \sum_{j=1}^{m_i}\int_0^t\omega_{ij}(u)\left[\bff\left\{\bfX_{ij}(u)\right\}\bbI\left\{\bfX_{ij}(u)\leq\bfx\right\}-\bfg(u,\bfx)\right]\rmd M_{ij}^k(\bfbeta_k,u) \nonumber\displaybreak[0]\\
	&-\bfh(t,\bfx)\bfA^{-1}(\tau)\sum_{j=1}^{m_i}\int_0^\tau\left\{\bfX_{ij}(t)-\tilde{\bfx}(t)\right\}\rmd M_{ij}^k(\bfbeta_k,t). 
\end{align*}
Furthermore, the covariance structure $\bbE\{\bfQ_i(t,\bfx)\bfQ_i(t^*,\bfx^*)'\}$ can be consistently estimated by $n^{-1}\sum_{i=1}^n\hat{\bfQ}_i(t,\bfx)\times\allowbreak\hat{\bfQ}_i(t^*,\bfx^*)'$, 
where
\begin{align*} 
	\hat{\bfQ}_i(t,\bfx) =& \sum_{j=1}^{m_i}\int_0^t\hat{\omega}_{ij}(u)\left[\bff\left\{\bfX_{ij}(u)\right\}\bbI\left\{\bfX_{ij}(u)\leq\bfx\right\}-\hat{\bfg}(u,\bfx)\right]\rmd \hat{M}_{ij}^k(\hat{\bfbeta}_k,u) \nonumber\displaybreak[0]\\
	&-\hat{\bfh}(t,\bfx)\hat{\bfA}^{-1}(\tau)\sum_{j=1}^{m_i}\int_0^\tau\left\{\bfX_{ij}(t)-\hat{\bfX}(t)\right\}\rmd \hat{M}_{ij}^k(\hat{\bfbeta}_k,t). 
\end{align*}

To operationalize the goodness-of-fit tests based on the cumulative sum of residuals with IPCW, we approximate the limiting distribution of $n^{-1/2}\bfW(t,\bfx)$ through a Monte Carlo simulation technique. Specifically, by repeatedly drawing simple random samples $\{\xi_1,\dots,\xi_n\}$ from the $\calN(0,1)$, we obtain the perturbed version of the weighted stochastic process
\begin{align} \label{eq:mc-w-hat}
	\hat{\bfW}(t,\bfx) = \sum_{i=1}^n\hat{\bfQ}_i(t,\bfx)\xi_i.
\end{align}
Goodness-of-fit tests can then be carried out using the limiting distribution approximated by the empirical distribution of the perturbed cumulative residual processes. The following result provides a theoretical justification for this perturbation procedure as a basis for model checking, with proof presented in Web Appendix A5.
\begin{theorem} \label{thm:mc-1}
	Given the observed data $\{N_{ij}^k(t),Y_{ij}^k(t),\bfX_{ij}(t),Z_{ij}(t),\allowbreak t\in[0,\tau],\allowbreak i=1,\dots,n,\allowbreak j=1,\dots,m_i\}$, the two stochastic processes, $n^{-1/2}\hat{\bfW}(t,\bfx)$ and $n^{-1/2}\bfW(t,\bfx)$, are asymptotically equivalent in $l^{\infty}[0,\tau]$, and both converge weakly to the same mean-zero Gaussian process, $n^{-1/2}\tilde{\bfW}(t,\bfx)$.
\end{theorem}

The cumulative sum of residuals $\hat{\bfW}(t,\bfx)$ can be utilized for checking different aspects of model specification with different specifications of $f(\cdot)$. 
For example, to assess the additive structure of the subdistribution hazards function, we consider the following weighted score-type process
\begin{align} \label{eq:mc-score}
	\bfU(\hat{\bfbeta}_k,t)=&\sum_{i=1}^n\sum_{j=1}^{m_i}\int_0^t \hat{\omega}_{ij}(u)\left\{\bfX_{ij}(u)-\hat{\bfX}(u)\right\}\left\{\rmd N_{ij}^k(u)-Y_{ij}^k(u)\bfX_{ij}(u)'\hat{\bfbeta}_k\rmd u\right\},
\end{align}
which is a special case of $\bfW(t,\bfx)$ with $\bff\{\bfX_{ij}(t)\}=\bfX_{ij}(t)$ and $\bfx=\bm{\infty}$. Using the Taylor's series expansion, we can show that
\begin{align*}
	n^{-1/2}\bfU(\hat{\bfbeta}_k,t)=&n^{-1/2}\bfU(\bfbeta_{k,0},t)-n^{-1/2}\bfA(t)(\hat{\bfbeta}_k-\bfbeta_{k,0})+\bm{o}_{\bm{p}}(\bm{1})\displaybreak[0]\\
	=&n^{-1/2}\sum_{i=1}^n\sum_{j=1}^{m_i}\left[\int_0^t\hat{\omega}_{ij}(u)\left\{\bfX_{ij}(u)-\hat{\bfX}(u)\right\}\rmd M_{ij}^k(\bfbeta_{k0},u)\right.\displaybreak[0]\\
	&\left.-\bfA(t)\bfA^{-1}(\tau)\int_0^\tau\hat{\omega}_{ij}(u)\left\{\bfX_{ij}(u)-\hat{\bfX}(u)\right\}\rmd M_{ij}^k(\bfbeta_{k0},u)\right]+\bm{o}_{\bm{p}}(\bm{1}).
\end{align*}
This result suggests a consistent covariance estimator for $n^{-1/2}\bfU(\bfbeta_{k0},\tau)$ is $\hat{\bfSigma}=n^{-1}\sum_{i=1}^n\hat{\bfPhi}_i(\hat{\bfbeta}_k,\tau)\hat{\bfPhi}_i(\hat{\bfbeta}_k,\tau)'$, 
where
\begin{align} \label{eq:mc-phi}
	\hat{\bfPhi}_i(\hat{\bfbeta}_k,t) = \sum_{j=1}^{m_i}\int_0^t\hat{\omega}_{ij}(u)\left\{\bfX_{ij}(u)-\hat{\bfX}(u)\right\}\rmd\hat{M}_{ij}^k(\hat{\bfbeta}_k,u).
\end{align}

Because the weighted-score type stochastic processes fluctuate randomly around the zero-axis under the null hypothesis of correct subdistribution hazards model specification, we construct the following goodness-of-fit tests from the maximum deviation of the processes from zero. Specifically, we write the test statistic for checking the additive structure of the $l$th covariate ($l=1,\dots,p$) in the subdistribution hazard as
\begin{align} \label{eq:mc-add-test-ind}
	\calS_l = \sup_{t\in[0,\tau]}\{\hat{\bfSigma}^{-1}\}_{ll}^{1/2}|n^{-1/2}U_l(\hat{\bfbeta}_k,t)|,
\end{align}
where $U_l(\hat{\bfbeta}_k,t)$ denotes the $l$th component of $\bfU(\hat{\bfbeta}_k,t)$ and $\{\hat{\bfSigma}^{-1}\}_{ll}$ denotes the $l$th diagonal element of $\hat{\bfSigma}^{-1}$. We write $s_l$ as the observed value of $\calS_l$ and  $\hat{\calS}_l=\sup_{t\in[0,\tau]}\{\hat{\bfSigma}^{-1}\}_{ll}^{1/2}|n^{-1/2}\hat{U}_l(\hat{\bfbeta}_k,t)|$, where $\hat{U}_l(\hat{\bfbeta}_k,t)$ is the $l$th component of the perturbed score process $\hat{\bfU}(\hat{\bfbeta}_k,t)$. The associated $p$-value for this test can be estimated by  
the proportion of $\hat{\calS}_l>s_l$ over the simulated distribution of $\hat{\calS}_l$ through repeated perturbation. Similarly, the test statistic for the joint additivity across all $p$ covariates can be considered as the sum of all $p$ individual statistics
\begin{align} \label{eq:mc-add-test-all}
	\calS_{\text{all}} = \sup_{t\in[0,\tau]}\sum_{l=1}^p\{\hat{\bfSigma}^{-1}\}_{ll}^{1/2}|n^{-1/2}U_l(\hat{\bfbeta}_k,t)|,
\end{align}
with the $p$-value estimated in a similar fashion through perturbation.

In order to assess the functional form of the $l$th covariate, $X_{ijl}(t)$, we can take $f_m\{\bfX_{ij}(t)\}=1$, $t=\tau$ and $\bfx_m=\infty$ for all $m\neq l, m = 1,\dots,p)$ and obtain the following form of the weighted cumulative residual process
\begin{align*}
	W_l(\tau,x_l) = \sum_{i=1}^n\sum_{j=1}^{m_i}\int_0^\tau \hat{\omega}_{ij}(t)\bbI\{X_{ijl}(t)\leq x_l\}\rmd\hat{M}_{ij}^k(\hat{\bfbeta}_k,t). 
\end{align*}
Similar to assessing the additive structure for each covariate, the null distribution of $W_l(\tau,x_l)$ (under the null that the functional form for the $l$th covariate is correct) can be approximated by the simulated mean-zero Gaussian process with perturbation. One can then obtain a $p$-value for the supremum test $\sup_{x_l}|W_l (\tau,x_l)|$ by generating a large number of realizations of $\hat{W}_l(\tau,x_l)$, where $\hat{W}_l(\tau,x_l)$ is the $l$th component of (\ref{eq:mc-w-hat}). 

To summarize, the test based on $U_l(\hat{\bfbeta}_k,t)$ is designed to identify potential non-additivity of $X_{ijl}(t)$ in the subdistribution hazard, whereas the test based on $W_l(\tau,x_l)$ is designed to identify a potentially incorrect functional form of $X_{ijl}(t)$ in the subdistribution hazard, under the assumption that $X_{ijl}(t)$ is independent of other covariates and that no other type of model misspecification exists. 
Furthermore, it is possible to show, similar to \citet{Lin1994} and \citet{Yin2007}, that these tests are consistent under the alternative hypotheses. In the Section \ref{sec:sim}, we demonstrate that our goodness-of-fit tests have power against specific types of model misspecification. 

\section{Simulation studies} \label{sec:sim}

We conducted two sets of simulation studies to assess the performance of our proposed methods. In the first simulation study, we validated our estimation strategy by comparing the results obtained from the CC data (as a gold standard) to those from the right-censored data via IPCW. We also demonstrated the necessity of accounting for clustering through a comparison with the approach developed in \citet{Li2017} for i.i.d. competing risks data. In the second simulation study, we examined the performance of our proposed goodness-of-fit tests, when no or certain aspects of the marginal additive subdistribution hazards model are misspecified. 

\subsection{Simulation study 1}\label{sec:sim1}
We simulated clustered competing risks data from the additive subdistribution hazards model with a primary event ($k=1$) and a competing event ($k=2$). 
We considered $\bfX_{ij}(t)=\bfX_{ij}e^{-t}$ as the set of time-varying covariates and ensured the CIF $F_k(\infty;\bfX)<1$, for $k=1,2$. 
Following the design in \citet{Fine1999} by defining $\rho=F_1(\infty;\bm{0})=\bbP\{\epsilon=1|\bfX_{ij}(t)=\bm{0}\}$ as the primary event rate in the reference group and $0<\rho<1$, we generated survival data in each cluster based on the following CIFs:
\begin{align*}
	F_1(t;\bfX_{ij},\nu_i) &= 1-\left\{1-(\rho+\nu_i)\left(1-e^{-t}\right)\right\}\exp\left\{-\bfX_{ij}'\bfbeta_1\left(1-e^{-t}\right)\right\}, \displaybreak[0]\\
	F_2(t;\bfX_{ij},\nu_i) &= \{1-(\rho+\nu_i)\}\exp\left(-\bfX_{ij}'\bfbeta_2\right)\left[1-\exp\left\{-t-\bfX_{ij}'\bfbeta_2\left(1-e^{-t}\right)\right\}\right],
\end{align*}
where $\nu_i$ is the cluster-specific frailty generated from a mean-zero exponential distribution of rate parameter $\theta$. Additional constraints are placed to ensure that $0<\rho+\nu_i<1$ and the CIFs are valid. While the above data generating process is based on cluster-level frailty $\nu_i$, we show in Web Appendix A6 that the marginal subdistribution hazards model structure of the primary event ($k=1$) still holds after integrating over the frailty distribution. 

We operationalized the following steps (note: for illustration, we consider a single covariate ($p=1$) such that $\beta_1$ is the scalar parameter of interest) to generate the observed data for all participants:  
\begin{itemize}
	\item[(i).] Specify cluster size $m_i$, for all $i$;
	\item[(ii).] Generate covariate $X_{ij}$ from a uniform distribution, $\calU(0,1)$, for all $i$ and $j$;
	\item[(iii).] Generate cluster-specific frailty $\nu_i$ from a demeaned exponential distribution with rate parameter $\theta$, for all $i$;
	\item[(iv).] Generate the cause of failure type $\epsilon_{ij}$, for all $i$ and $j$:
	\begin{itemize}
		\item[(a).] Compute the probability for competing risk 1:
		\begin{align*}
			P_{ij1}=F_1(\infty;X_{ij},\nu_i) = 1-\{1-(\rho+\nu_i)\}\exp\left(-X_{ij}\beta_1\right);
		\end{align*}
		\item[(b).] Generate $U_{ij1}$ from $\calU(0,1)$;
		\item[(c).] Generate cause of failure type $\epsilon_{ij}$:
		\begin{align*}
			\epsilon_{ij}=\left\{\begin{array}{cc}
				1,& \text{if}~ U_{ij1}\leq P_{ij1}  \\
				2,& \text{if}~ U_{ij1}> P_{ij1}  
			\end{array}\right.;
		\end{align*}
	\end{itemize}        
	\item[(v).] Generate failure time $T_{ij}$ from the conditional distribution
	\begin{align*}
		\tilde{F}_k(t;X_{ij},\nu_i) = \frac{\bbP(T_{ij}\leq t,\epsilon_{ij}=k|X_{ij},\nu_i)}{\bbP(\epsilon_{ij}=k|X_{ij},\nu_i)} = \frac{F_k(t;X_{ij},\nu_i)}{F_k(\infty;X_{ij},\nu_i)},
	\end{align*}
	based on $\epsilon_{ij}$, using the inverse distribution method;
	\item[(vi).] Generate censoring time $C_{ij}$ from a pre-specified censoring time distribution and let $\zeta_{ij}=\bbI(T_{ij}\leq C_{ij})\epsilon_{ij}$, for all $i$ and $j$.
\end{itemize}

A variety of parameter configurations were considered to examine the performance of the proposed method. We specified cluster sizes of 10 and 20 and numbers of clusters of 100 and 250. Values for $\rho$, $\beta_1$, and $\beta_2$ were set as 0.5, 1.0, and 0.2, respectively. We assumed that $\beta_1$ was parameter of interest and $\beta_2$ was the regression parameter in the CIF of the competing event. The rate parameter $\theta$, which determines the distribution of cluster-specific frailty, took values 0.7 and 1.0. For simplicity, we generated independent censoring times from the exponential distribution with rate parameter $\gamma\in\{0.35,0.95\}$, such that the marginal censoring proportion was around $20\%$ and $40\%$, respectively. For each parameter combination, we simulated 1000 replicates and compared the proposed method (C) and the unclustered additive subdistribution hazards model (UC) developed in \citet{Li2017} for the purpose of estimating $\beta_1$. We also created two types of data,  complete censoring (CC) data and right-censored (RC) data, and considered both methods for each data type. The simulation results under the CC data represent the gold standard and were used to assess the effectiveness of IPCW under RC data in recovering the potentially unobserved censoring information in practical applications.

Table \ref{tab:sim-1-1} and Table \ref{tab:sim-1-2} summarize the simulation results under two censoring proportions, including the average point estimates ($\bbE(\hat{\beta}_1)$), Monte Carlo standard error (MCSE, $s(\hat{\beta}_1)$), average of the estimated standard error (AESE, $\bbE(\hat{s})$) and 95\% confidence interval coverage. 
The average parameter estimates remain close for all approaches under all settings, which is expected because our approach assumes a working independence correlation assumption and therefore considers the same (weighted) least squares estimator for $\beta_1$ as in the unclustered approach. However, the two clustered approaches (CRC, CCC) have better performance compared to the unclustered approaches (UCRC, UCCC) in terms of the agreement between the MCSE and AESE, and the coverage of the confidence intervals. This result implies that ignoring the clustering structure in the data may lead to an overly narrow confidence interval estimate and an inflated type I error rate under the null hypothesis (e.g., $H_0:\beta_1=1$). Furthermore, bias in estimating the variance and hence under-coverage due to ignoring the clustering structure appears to be more pronounced as cluster size increases from 10 to 20. 
Finally, when the data are RC, our proposed method produces results that are almost identical to those when the data are CC (gold standard), with accurate variance estimates and nominal coverage under both censoring proportions.

\begin{table}[htbp] 
	\caption{Simulation results when the marginal censoring proportion is around $20\%$. The true value of $\beta_1$ is $1$ and $\theta$ refers to the rate parameter of the frailty distribution in the data generating process. CRC: clustered and right-censored; CCC: clustered and complete-censoring; UCRC: unclustered and right-censored; UCCC: unclustered and complete-censoring.}
	\label{tab:sim-1-1}\par
	\centering
	\resizebox{0.67\linewidth}{!}{
		\begin{tabular}{ccccccc} \hline 
			\multicolumn{7}{c}{Number of Clusters = 100}\\ \hline
			Cluster Size & $\theta$ & Approaches & $\bbE(\hat{\beta}_1)$ & $s(\hat{\beta}_1)$ & $\bbE(\hat{s})$ & Coverage (\%) \\ \hline
			10 & 0.7 & CRC & 1.016 & 0.225 & 0.233 & 95.20  \\
			& & CCC & 1.017 & 0.227 & 0.233 & 95.20  \\
			& & UCRC & 1.016 & 0.225 & 0.192 & 88.40  \\
			& & UCCC & 1.017 & 0.227 & 0.192 & 88.10  \\
			& 1.0 & CRC & 1.024 & 0.246 & 0.231 & 94.40  \\
			& & CCC & 1.024 & 0.248 & 0.231 & 94.30 \\
			& & UCRC & 1.024 & 0.246 & 0.189 & 87.60 \\
			& & UCCC & 1.024 & 0.248 & 0.189 & 87.70  \\
			20 & 0.7 & CRC & 1.005 & 0.192 & 0.184 & 94.30 \\
			& & CCC & 1.005 & 0.192 & 0.184 & 94.20 \\
			& & UCRC & 1.005 & 0.192 & 0.136 & 87.40 \\
			& & UCCC & 1.005 & 0.192 & 0.136 & 87.20 \\
			& 1.0 & CRC & 1.009 & 0.206 & 0.199 & 94.30 \\
			& & CCC & 1.009 & 0.207 & 0.199 & 94.30 \\
			& & UCRC & 1.009 & 0.206 & 0.138 & 87.00 \\
			& & UCCC & 1.009 & 0.207 & 0.139 & 86.60 \\ \hline
			\multicolumn{7}{c}{Number of Clusters = 250} \\ \hline
			Cluster Size & $\theta$ & Approaches & $\bbE(\hat{\beta}_1)$ & $s(\hat{\beta}_1)$ & $\bbE(\hat{s})$ & Coverage (\%) \\ \hline
			10 & 0.7 & CRC & 0.995 & 0.145 & 0.146 & 96.10 \\
			& & CCC & 0.995 & 0.146 & 0.146 & 96.00 \\
			& & UCRC & 0.995 & 0.145 & 0.121 & 89.50 \\
			& & UCCC & 0.995 & 0.146 & 0.121 & 89.50 \\
			& 1.0 & CRC & 0.987 & 0.152 & 0.144 & 94.10 \\
			& & CCC & 0.988 & 0.153 & 0.145 & 94.30 \\
			& & UCRC & 0.987 & 0.152 & 0.124 & 86.90 \\
			& & UCCC & 0.988 & 0.153 & 0.124 & 87.20 \\
			20 & 0.7 & CRC & 0.998 & 0.122 & 0.121 & 95.50 \\
			& & CCC & 0.998 & 0.122 & 0.122 & 95.40 \\
			& & UCRC & 0.998 & 0.122 & 0.089 & 87.70 \\
			& & UCCC & 0.998 & 0.122 & 0.089 & 87.60 \\
			& 1.0 & CRC & 1.002 & 0.126 & 0.124 & 94.40 \\
			& & CCC & 1.002 & 0.126 & 0.124 & 94.30 \\
			& & UCRC & 1.002 & 0.126 & 0.091 & 88.30 \\
			& & UCCC & 1.002 & 0.126 & 0.092 & 87.60
			\\ \hline
		\end{tabular}
	}
\end{table}

\begin{table}[htbp] 
	\caption{Simulation results when the marginal censoring rate is around $40\%$. The true value of $\beta_1$ is $1$ and $\theta$ refers to the rate parameter of the frailty distribution in the data generating process. CRC: clustered and right-censored; CCC: clustered and complete-censoring; UCRC: unclustered and right-censored; UCCC: unclustered and complete-censoring.}
	\label{tab:sim-1-2}\par
	\centering
	\resizebox{0.67\linewidth}{!}{
		\begin{tabular}{ccccccc} \hline 
			\multicolumn{7}{c}{Number of Clusters = 100}\\ \hline
			Cluster Size & $\theta$ & Approaches & $\bbE(\hat{\beta}_1)$ & $s(\hat{\beta}_1)$ & $\bbE(\hat{s})$ & Coverage (\%) \\ \hline
			10 & 0.7 & CRC & 0.986 & 0.247 & 0.244 & 94.60 \\
			& & CCC & 0.989 & 0.247 & 0.244 & 94.60 \\
			& & UCRC & 0.986 & 0.247 & 0.213 & 90.60 \\
			& & UCCC & 0.989 & 0.247 & 0.214 & 90.10 \\
			& 1.0 & CRC & 0.999 & 0.243 & 0.232 & 95.80 \\
			& & CCC & 1.001 & 0.244 & 0.232 & 95.70 \\
			& & UCRC & 0.999 & 0.243 & 0.208 & 89.70 \\
			& & UCCC & 1.001 & 0.244 & 0.209 & 90.10 \\
			20 & 0.7 & CRC & 0.994 & 0.209 & 0.194 & 94.10 \\
			& & CCC & 0.994 & 0.209 & 0.194 & 94.20 \\
			& & UCRC & 0.994 & 0.209 & 0.155 & 87.80 \\
			& & UCCC & 0.994 & 0.209 & 0.155 & 87.20 \\
			& 1.0 & CRC & 1.005 & 0.202 & 0.198 & 95.80 \\
			& & CCC & 1.005 & 0.202 & 0.199 & 95.80 \\
			& & UCRC & 1.005 & 0.202 & 0.146 & 88.60 \\
			& & UCCC & 1.005 & 0.202 & 0.147 & 88.80 \\ \hline
			\multicolumn{7}{c}{Number of Clusters = 250} \\ \hline
			Cluster Size & $\theta$ & Approaches & $\bbE(\hat{\beta}_1)$ & $s(\hat{\beta}_1)$ & $\bbE(\hat{s})$ & Coverage (\%) \\ \hline
			10 & 0.7 & CRC & 1.004 & 0.156 & 0.151 & 95.50 \\
			& & CCC & 1.004 & 0.157 & 0.151 & 95.30 \\
			& & UCRC & 1.004 & 0.156 & 0.137 & 90.40 \\
			& & UCCC & 1.004 & 0.157 & 0.137 & 90.50 \\
			& 1.0 & CRC & 1.008 & 0.154 & 0.152 & 96.40 \\
			& & CCC & 1.008 & 0.154 & 0.152 & 96.40 \\
			& & UCRC & 1.008 & 0.154 & 0.130 & 89.90 \\
			& & UCCC & 1.008 & 0.154 & 0.130 & 90.20 \\
			20 & 0.7 & CRC & 1.001 & 0.133 & 0.131 & 94.80 \\
			& & CCC & 1.001 & 0.132 & 0.131 & 94.60 \\
			& & UCRC & 1.001 & 0.133 & 0.096 & 88.70 \\
			& & UCCC & 1.001 & 0.132 & 0.096 & 88.80 \\
			& 1.0 & CRC & 1.006 & 0.125 & 0.123 & 95.50 \\
			& & CCC & 1.006 & 0.125 & 0.123 & 95.60 \\
			& & UCRC & 1.006 & 0.125 & 0.096 & 89.40 \\
			& & UCCC & 1.006 & 0.125 & 0.097 & 88.90
			\\ \hline
		\end{tabular}
	}
\end{table}

\subsection{Simulation study 2}

We evaluated the performance of the proposed model checking procedure in examining the additive structure in the subdistribution hazards model. For illustration, we considered the test for assessing the overall model fit based on the test statistic \eqref{eq:mc-add-test-all}. To assess the validity of the goodness-of-fit tests under the correct model specification, we simulated clustered competing risks data from the following marginal additive subdistribution hazards model (model $M_1$):
\begin{align*}
	F_{M_11}\left(t;\bfX_{ij},\nu_i\right) &= 1-\left\{1-(\rho+\nu_i)\left(1-e^{-t}\right)\right\}\exp\left\{-\bfX_{ij}'\bfbeta_{M_11}\left(1-e^{-t}\right)\right\}, \displaybreak[0]\\
	F_{M_12}\left(t;\bfX_{ij},\nu_i\right) &= \{1-(\rho+\nu_i)\}\exp\left(-\bfX_{ij}'\bfbeta_{M_12}\right)\left[1-\exp\left\{-t-\bfX_{ij}'\bfbeta_{M_12}\left(1-e^{-t}\right)\right\}\right],
\end{align*}
where $\rho$ and $\nu_i$ are defined as in Section \ref{sec:sim1}. The two CIFs in model $M_1$, $F_{M_11}(t;\bfX,\nu)$ and $F_{M_12}(t;\bfX,\nu)$, have the same functional structure as their corresponding counterparts, $F_1(t;\bfX,\nu)$ and $F_2(t;\bfX,\nu)$, in simulation study 1. We added subscripts to CIFs and their related parameters, $\bfbeta_{M_1k}$ and $\bfbeta_{M_2k}$ ($k=1,2$), simply to distinguish between involved models in simulation study 2. The goodness-of-fit test is expected to exhibit empirical type I error rates around the nominal 0.05 level when applied to data generated from model $M_1$, since the CIF of the primary event, $F_{M_11}(t;\bfX,\nu)$ follows an additive structure and will be correctly specified.

To assess the power of the test when the additive subdistribution hazards model is misspecified, we simulated clustered competing risks data from the marginal proportional subdistribution hazards model (model $M_2$), defined as
\begin{align*}
	F_{M_21}\left(t;\bfX_{ij},\nu_i\right) &= 1-\left\{1-(\rho+\nu_i)\left(1-e^{-t}\right)\right\}^{\exp\left\{-\bfX_{ij}'\bfbeta_{M_21}\left(1-e^{-t}\right)\right\}}, \\
	F_{M_22}\left(t;\bfX_{ij},\nu_i\right) &= \{1-(\rho+\nu_i)\}^{\exp\left(-\bfX_{ij}'\bfbeta_{M_22}\right)}\left[1-\exp\left\{-t\bfX_{ij}'\bfbeta_{M_22}\left(1-e^{-t}\right)\right\}\right].
\end{align*}
Here, we considered two covariates ($p=2$) in both data generating processes, where $X_{ij1}\sim \calN(0,1)$, and $X_{ij2}$ was generated from a Bernoulli distribution with $\bbP(X_{ij2}=1)=\bbP(X_{ij2}=0)=0.5$. We set $\bfbeta_{M_11}=(0.6,1)'$, $\bfbeta_{M_21}=(0.5,1)'$, $\bfbeta_{M_12}=\bfbeta_{M_22}=(0.5,1)'$, and the primary event rate among the reference group, $\rho=0.66$. The type I error rate and the power were evaluated under different settings by varying the distribution of the cluster-specific frailty, censoring percentage, and number of clusters similar to Section \ref{sec:sim1}. Specifically, we used the mean-zero exponential distribution with rate parameter 0.7 and 1.0 to generate the cluster-specific frailty, as well as an exponential distribution with rate parameter $\in\{0.35, 0.95, 1.65\}$ to generate censoring times, corresponding to 20\%, 40\%, and 60\% censoring proportions, respectively. For simplicity, the cluster size was set as 10, and number of clusters given as 100 and 150. For each scenario, we simulated 1000 replicates to evaluate the type I error rate and power of the goodness-of-fit test. The null distribution of the test statistic for each data replication was approximated by randomly drawing 1000 perturbed stochastic processes. 

As shown in Table \ref{tab:sim-2}, the proposed overall goodness-of-fit test is valid in that the associated empirical type I error rates are close to the nominal value of 0.05. The test also shows adequate power to reject the null of no model misspecification, especially when the number of clusters increases and the censoring proportion decreases. This trend is expected because the power of the test will depend on the amount of observed information in the data to support diagnosis of lack of fit. Furthermore, a larger variance of the frailty or a higher degree of within-cluster correlation of the failure times can also lead to a slight decrease in the power of the goodness-of-fit test.

\begin{table}[h!] 
	\caption{Empirical type I error rates and power of the overall goodness-of-fit test when the clustered competing risks data are generated under model $M_1$ (when the null holds) and $M_2$ (when the alternative holds), respectively. The parameter $\theta$ refers to the the rate parameter of the frailty distribution in the data generating process. The cluster size is fixed at 10 with a nominal significance level $\alpha=0.05$.}
	\label{tab:sim-2}\par
	\centering
		\begin{tabular}{cccccc} \hline 
			Number & Censoring & \multicolumn{2}{c}{$\theta=0.7$} & \multicolumn{2}{c}{$\theta=1.0$} \\ \cline{3-6}
			of Clusters & (\%) & Type I error ($M_1$) & Power ($M_2$) & Type I error ($M_1$) & Power ($M_2$)\\\hline
			100 & 20 & 0.047 & 0.678 & 0.051 & 0.662 \\
			& 40 & 0.063 & 0.384 & 0.064 & 0.423 \\
			& 60 & 0.072 & 0.156 & 0.069 & 0.153 \\
			150 & 20 & 0.052 & 0.966 & 0.058 & 0.894 \\
			& 40 & 0.061 & 0.658 & 0.057 & 0.672 \\
			& 60 & 0.068 & 0.371 & 0.063 & 0.254
			\\ \hline
		\end{tabular}
\end{table}

\section{Analysis of the STRIDE study} \label{sec:app}

The STrategies to Reduce Injuries and Develop confidence in Elders (STRIDE) study is a pragmatic cluster randomized trial focusing on reducing serious fall injuries in community-dwelling older adults at risk of falls \citep{bhasin_strategies_2018,bhasin_nejm_2020}. The study enrolled 5451 adults aged 70 and older from 86 primary care practices; each practice was randomized to either an evidence-based fall-related injury prevention program or enhanced usual care in a 1:1 ratio. The event of interest was time to first serious fall-related injury, and death without fall injury was considered as a competing cause of failure. The rates of fall-related injury among the control and intervention practices were approximately 5.3 and 4.9 per 100 person-years of follow-up, while the observed competing event rate was 3.3 per 100 person-years of follow-up in both intervention and control practices \citep{bhasin_nejm_2020}. Patients withdrew consent at a rate of 3.6 per 100 person-years of follow-up and are right-censored; 4187 participants were administratively censored.

Since this is a cluster randomized trial with a heterogeneous patient population across primary care practices, there may be unobserved factors that are shared across patients in each practice, and failure to account for clustering may result in invalid inference \citep{turner2017review}. We consider the marginal additive subdistribution hazards model \eqref{eq:model-def} with the proposed robust sandwich variance estimator. In addition to the intervention indicator, we consider location of the practice (urban vs rural), age, gender, race (white vs non-white) and number of chronic coexisting conditions as five potential risk factors to adjust for in the model. In particular, location of the practice and race were balanced in the design stage via constrained randomization and necessitates adjustment to maintain valid inference \citep{li2017evaluation}.

\begin{figure}[htbp]
	\centering
	\includegraphics[width=\textwidth]{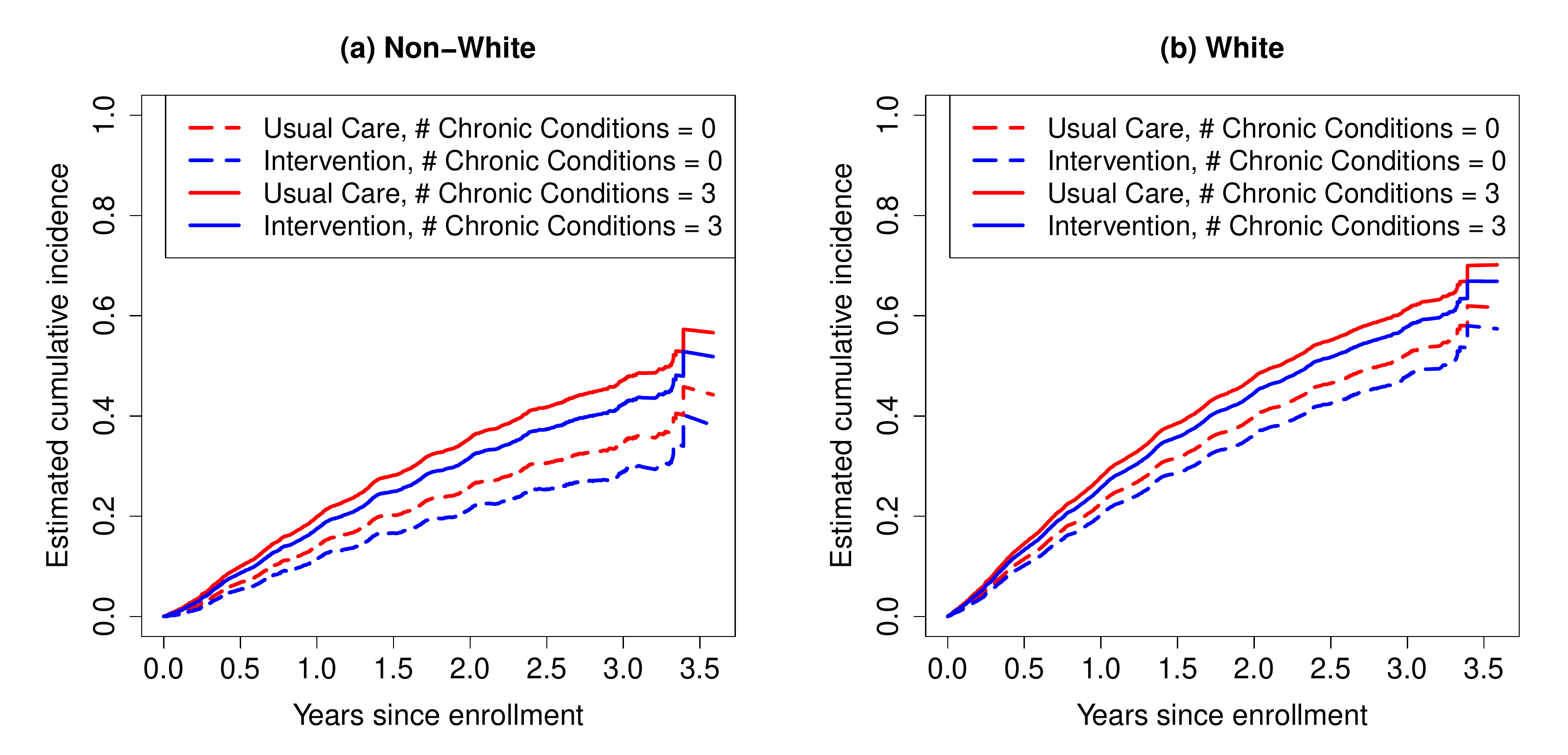}
	\caption{{Estimated cumulative incidence functions for self-reported fall injury among four typical White and non-White patients. Each patient is assumed to come from an urban practice, with age 76 years old and female.}}
	\label{fig:CIFplot}
\end{figure}

Table \ref{tab:app-1} presents the estimated coefficients and robust standard errors from our marginal model. Although, none of the effects reach statistical significance at the 0.05 level (as the absolute value of each estimate over the robust standard error does not exceed 1.96), the intervention effect appears to favor the fall-related injury prevention program ($\hat{\beta}_1=-0.0292$), with those receiving the intervention being at slightly lower risk of a fall related injury. Among other risk factors, female and white patients from an urban practice tend to have a larger risk for falls in the absence of intervention and the number of chronic coexisting conditions further increases the risk for falls. In particular, we found older age to be associated with a decreasing risk for falls ($\hat{\beta}_3=-0.0009$). This negative association is likely because the recruited population are at least 70 years old, among whom a further increase in age could start to prevent them from potential triggers for serious falls such as exercise or intensive movement. To further illustrate our method, we present the estimated cumulative incidence function for eight hypothetical patients in Figure \ref{fig:CIFplot}. Here we consider each patient as a 76-year-old (mode of the study population), female, seeking care from an urban practice, and present the combinations of intervention vs usual care, white vs non-white and no chronic coexisting conditions vs 3 chronic coexisting conditions. The figure clearly demonstrates that white patients are at a higher risk for falls than non-white patients, and the intervention program reduces the risk for falls and with a larger magnitude of absolute risk reduction over time.

\begin{table}[h!] 
	\caption{Estimation and model checking results for the analysis of STRIDE study.}
	\label{tab:app-1}\par
	\centering
		\begin{tabular}{rrccc} \hline 
			& \multicolumn{2}{c}{Model Fitting} & \multicolumn{2}{c}{Model Checking} \\ \cline{2-5}
			& Estimate & Robust SE & Test Statistic & $p$-Value\\ \hline
			\texttt{Intervention} ($\hat{\beta}_1$) & $-0.0292$ & $0.9148$ & $1.6645$ & $0.766$ \\
			\texttt{Urban} ($\hat{\beta}_2$) & $0.0311$ & $1.1603$ & $1.4319$ & $0.882$\\
			\texttt{Age} ($\hat{\beta}_3$) & $-0.0009$ & $0.0792$ & $0.8260$ & $0.847$\\
			\texttt{Female} ($\hat{\beta}_4$) & $0.0078$ & $0.9681$ & $0.8893$ & $0.668$ \\
			\texttt{White} ($\hat{\beta}_5$) & $0.1042$ & $1.2875$ & $0.8307$ & $0.892$ \\
			\texttt{\# Chronic Conditions} ($\hat{\beta}_6$) & $0.0234$ & $0.3693$ & $1.0146$ & $0.659$\\
			{Overall} & -- & -- & $4.5546$ & $0.998$ \\
			\hline
		\end{tabular}
\end{table}

\begin{figure}[htbp]
	\centering
	\includegraphics[width=1\textwidth]{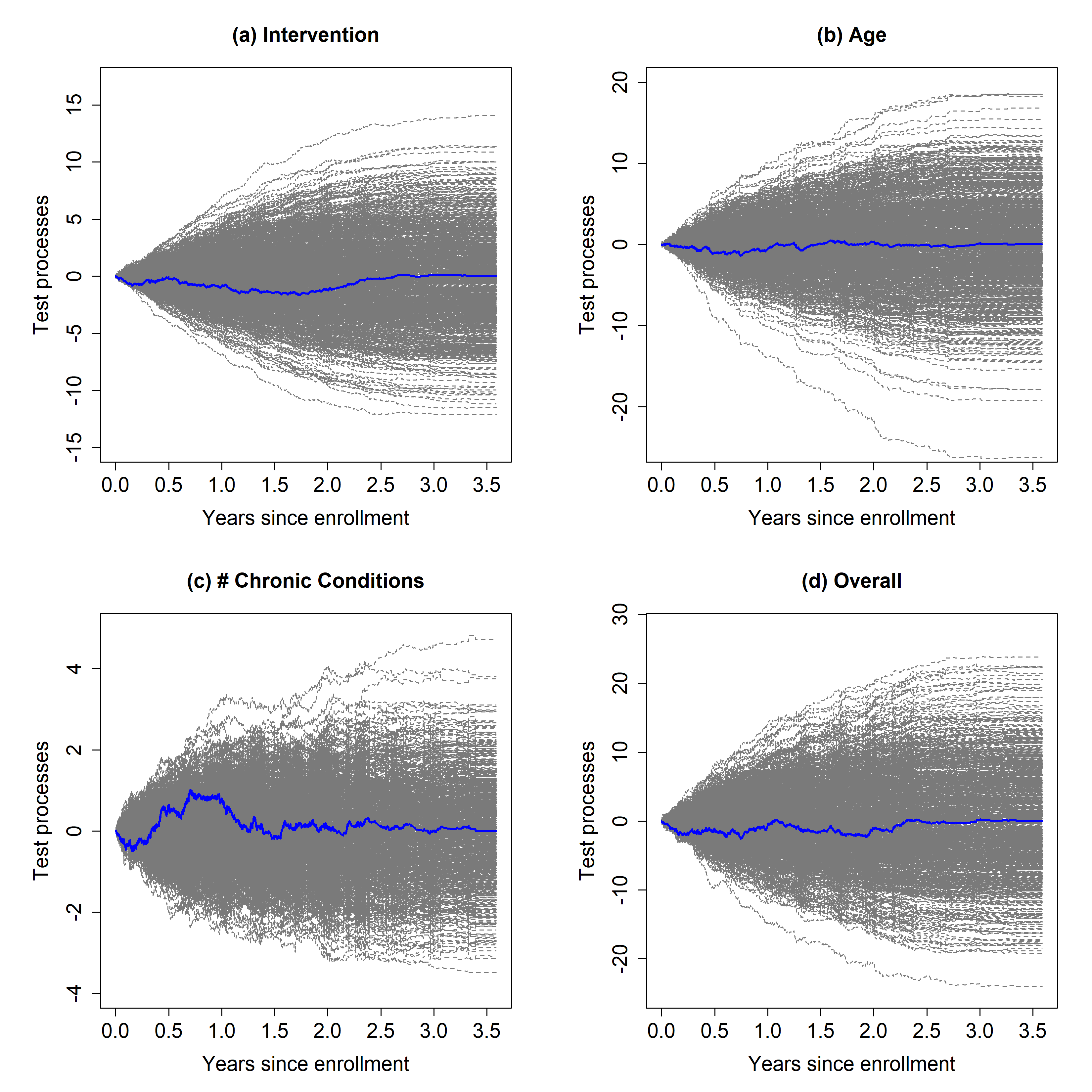}
	\caption{{Plot of the observed test process (blue) and simulated curves (gray) under the null in the STRIDE study for model checking based on three covariates (intervention, age and number of chronic coexisting conditions), as well as the overall fit.}}
	\label{fig:modelchecking}
\end{figure}

We examine the adequacy of the assumed additive structure for the subdistribution hazards function, by carrying out the proposed goodness-of-fit tests with results for each covariate, as well as the overall model fit. Each test is based on 1000 simulated test processes with the test statistics and $p$-values given in Table \ref{tab:app-1}. The $p$-value for each covariate is at least 0.659, which supports the appropriateness of the additive structure assumed for each covariate. The test for overall model fit yields a $p$-value of 0.998, indicating no evidence from the data against the additive assumption across all covariates in the marginal subdistribution hazards model. For four different tests, Figure \ref{fig:modelchecking} graphically illustrates that the observed test process can be completely covered by the 1000 simulated processes, suggesting no lack of fit. 

\section{Discussion} \label{sec:dis}

In this article, we developed the marginal additive subdistribution hazards model to analyze clustered survival data subject to competing risks data which provides a complementary approach to the marginal proportional subdistribution hazards model. Assuming working independence, the estimation was based on inverse probability of censoring weighted least squares and the theoretical properties of the estimator were studied. We provided a new robust sandwich variance to quantify the uncertainty of the regression estimators. Our simulations demonstrated the necessity of accounting for clustering through our new sandwich variance estimator to achieve nominal coverage probability in a range of realistic parameter configurations. To support the analysis of clustered competing risks data via the marginal additive subdistribution hazards model, we also developed a set of model checking procedures. Our simulations indicated that the proposed goodness-of-fit tests carry the nominal type I error rate and have sufficient power to detect aspects of model misspecification. 

There are several limitations that we plan to pursue in future work. First, while we offer a complementary approach for clustered competing risks regression without invoking the proportional subdistribution hazards assumption, we have not compared our approach with the marginal subdistribution hazards model for predicting the cumulative incidence function under different data generating processes to assess their robustness and relative efficiency. Second, we have assumed the working independence correlation model and regarded the complex within-cluster correlation structure as a nuisance parameter. While the robust sandwich variance can provide valid inference under working independence, a potential improvement of the marginal subdistribution hazards model may be made by further incorporating a precision weighting matrix in \eqref{eq:rc-score}. For example, an appropriate precision weighting matrix based on the pairwise martingale covariance process has been previously developed for more efficient estimation of the marginal Cox model in the absence of competing risks \citep{prentice1992covariance,cai1995estimating} and may be extended for more efficient parameter estimation in our model. Finally, we have assumed the censoring times are independent of the failure times and covariates in each cluster and used the KM estimator to estimate the inverse probability of censoring weights. It would be of interest to extend our approach under covariate-dependent censoring \citep{he2016proportional}, where the weights are computed from a marginal Cox model adjusting for covariates. 

\section*{Acknowledgements}\label{acknowledgements}

This work was partially supported by National Institutes of Health (NIH) grants UL1 TR0001863. The authors thank Drs. Peter Peduzzi and Erich Greene for their help in providing the data from the STrategies to Reduce Injuries and Develop confidence in Elders (STRIDE) study for our illustrative application. The STRIDE study was funded by the Patient Centered Outcomes Research Institute (PCORI), with additional support from the National Institute on Aging at NIH (U01AG048270) and the Claude D. Pepper Older Americans Independence Center at Yale School of Medicine (P30AG021342). The statements presented in this article are solely the responsibility of the authors and do not necessarily represent the views of NIH, PCORI\textsuperscript{\textregistered} or its Board of Governors or Methodology Committee. The authors are grateful to the editor and the two anonymous referees for their constructive suggestions, which have improved an older version of this paper.

\section*{Conflict of Interest}

The authors declare that they have no conflict of interest.

\bibliography{CCRR_MAM}

\end{document}